\documentclass[showpacs,twocolumn,pra,superscriptaddress]{revtex4}
\usepackage{amsmath}
\usepackage{graphicx}
\usepackage{dcolumn}
\usepackage{stackrel,amssymb}
\usepackage{color}
\usepackage{ulem}

\usepackage{bm}

\newcommand{\ket}[1]{|#1\rangle}

\begin{document}
\title{Quantum Darwinism and non-Markovian dissipative dynamics from quantum
phases of the spin$-1/2$ XX model}

\author{Gian Luca Giorgi}
\affiliation{Istituto Nazionale di Ricerca Metrologica, Strada delle Cacce 91
I-10135 Torino, Italy
}

\author{Fernando Galve}
\affiliation{Instituto de F\'{i}sica Interdisciplinar y Sistemas Complejos IFISC
(CSIC-UIB),
Campus Universitat Illes Balears, E-07122 Palma de Mallorca, Spain}

\author{Roberta Zambrini} 
\affiliation{Instituto de F\'{i}sica Interdisciplinar y Sistemas Complejos IFISC
(CSIC-UIB),
Campus Universitat Illes Balears, E-07122 Palma de Mallorca, Spain}


\begin{abstract}
Quantum Darwinism explains the emergence of a classical description of objects
in terms of the creation of many redundant
registers in an environment containing their classical information. This
amplification phenomenon, where only classical information
reaches the macroscopic observer and through which different observers can agree
on the objective existence of such object, has been revived
lately for several types of situations, successfully explaining classicality. We
explore quantum Darwinism in the setting of an environment made
of two level systems which are initially prepared in the ground state of the XX
model, which exhibits different phases; we find that the different
phases have different ability to redundantly acquire classical information about
the system, being the ``ferromagnetic  phase'' the only one able
to complete quantum Darwinism. At the same time we relate this ability to how
non-Markovian the system dynamics is, based on the interpretation that
non-Markovian dynamics is associated to back flow of information from
environment to system, thus spoiling the information transfer needed for
Darwinism.
Finally, we explore mixing of bath registers by allowing a small interaction
among them, finding that this spoils the stored information as previously found
in the literature.
\end{abstract}
\maketitle

\section{Introduction}\label{intro}

Quantum Darwinism deals with the description and the quantification of the
reasons why quasi-classical states play a special role in our world, 
as they are perfectly suitable to describe what happens  in a broad range of
scales \cite{zurekNAT}. Indeed, in our daily experience, we only 
observe phenomena that can be described by classical laws, despite the fact that
classical states represent a very tiny fraction of the whole Hilbert space of
the Universe. 

The theory of open quantum systems explains
the emergence of
classicality as the result of the interaction of a small system with a 
larger one representing the environment. Decoherence  is the tool used to
describe the transition from quantum to classical states through the 
loss of information from the system towards the environment \cite{deco1,deco2}.
When describing the emergence of decoherence, a partial trace is 
performed over the bath's degrees of freedom, that is, the information contained
in the environment is not accessible at all. Because of 
the decoherence process, there are states that naturally emerge due to their
stability with respect to the interaction with the bath. These states, 
usually referred to as  pointer states \cite{pointer}, are the best candidates to describe the
classical world. 

In addition, most of the information we have about any system comes from indirect
observations made on fragments of the environment rather than on the 
system itself. The fact that different observers accessing different fragments
get the same information about the system is  the central result of 
quantum Darwinism: the states perceived in the same way (objectively) by multiples observers
 are the ones that are able to spread
around them multiple copies of their classical information content. This
illustrates the emergence of an objective classical reality from the quantum
probabilistic world.  

Recently it has been shown \cite{brandao} that observers monitoring a system by
measuring its environment can only learn about a unique (pointer) observable, this being
a generic feature of quantum mechanics. However this is only part of the program:
observers need to be able to obtain close to full classical information on the 
system and agree among them. 
 This latter redundant proliferation of
information regarding pointer states,  far from being generic, has been demonstrated to take 
place in different physical contexts, that is,  
considering purely dephasing Hamiltonians \cite{zurekCHER}, photon environments
\cite{zurekPHOTON,korbicz}, spin environments \cite{spins1,haziness,zurekFALL}
and Brownian motion \cite{brownian1,brownian2,brownian3}. It has recently been stressed that this redundant encoding should be
checked at the level of states, with a spectrum broadcast structure \cite{origin}. 

The achievement of quantum Darwinism,  obviously  related to which
Hamiltonians govern the problem, is also determined by the initial state of the bath. 
The inhibition of redundancy in
Brownian motion has recently been linked \cite{galve14} to the presence of 
non-Markovianity in the open quantum system dynamics \cite{nm,nm2}. 
Intuitively: the rollback of the decoherence process, through which
the environment learns about the system, and which is a salient feature of
non-Markovian evolution, is expected to spoil records of the system 
imprinted upon the environment.
In the case of spins, it is known that mixedness and misalignment 
\cite{haziness}(the closeness of bath states to eigenstates of the interaction Hamiltonian)
of environmental units will reduce the bath's ability to produce Darwinism.
Also, if the bath units are interacting \cite{zurekFALL}, the redundant classical records
will spread and become inaccessible locally, which in the end forces observers
to collect huge amounts of bath fragments thus ruining Darwinism.

In this paper we will first consider the role of initial correlations in a separable
environment: an ensemble of (uncoupled) spins is prepared in the ground state of the (coupled)
XX model in the presence of an external magnetic field. In fact, by tuning the value of
the field, different ground states with different correlation 
properties are available. It will be shown that an initially uncorrelated environment
is more liable to produce quantum Darwinism.
Second, by gently turning on the XX coupling Hamiltonian among bath spins, we will show
that redundant classical records in the bath are spoiled in proportion to the coupling.
Finally, we analyze the non-Markovianity of the system's evolution showing that its 
behaviour follows the same trend of quantum Darwinism. We end our work conjecturing
a possible relation between the two phenomena, based on the amount of inter-correlations
present in the environment as a pernicious influence.

\section{Quantum Darwinism}

Quantum Darwinism is complete when the environment is able to store redundantly
copies of the classical information about a pointer observable of the system,
meaning
that any other information about the system has not `survived' the (time)
evolution. 
This is typically quantified by the mutual information between the system and
fractions ${\mathcal F}$ of the environment: classical, objective existence of
the system needs different
observers, who can access different and independent fractions of the
environment, to agree on the properties of the system by querying such
fractions. The size of the fractions 
should not be a limitation, since otherwise we would need a minimum amount of
environment to learn something about the system. Mathematically this condition
basically means that
the mutual information between system and fractions of the environment
$I(S:{\mathcal F})$ must be almost independent of fraction size ($\#\mathcal{F}$) and on which
fraction we have chosen
(i.e.  $I(S:\mathcal{F})\neq f(\#\mathcal{F})$ and $I(S:{\mathcal F}_1)=I(S:{\mathcal F}_2)$ when $\#\mathcal{F}_1=\#\mathcal{F}_2$). 

Let us consider a system $S$ in contact with an environment ${\cal E}$ and 
suppose that we have knowledge about the inner structure of ${\cal E}$, that is 
${\cal E}=\otimes_{i=1}^N {\cal E}_i$, where  each ${\cal E}_i$ has dimension
$D$. We also suppose that any individual  fragment of the environment 
${\cal F}=\otimes_{i=1}^{\#\mathcal{F}} {\cal E}_i$  \footnote{ Note that since the environmental
states are unchanged by permutation of its units, we can label any fragment with spin indices $1,2,...,fN$ without loss
of generality.}
of size $\#\mathcal{F}=f N$ ($f\cdot N$, with $0\leq f\leq 1$, is the  number of
${\cal E}$ 's spins  contained in ${\cal F}$) is accessible. The mutual information 
between $S$ and ${\cal F}$, which quantifies how much information about $S$ is
present in the fragment ${\cal F}$ is defined as
\begin{equation}
{\cal I}(S:{\mathcal F} )=H_S+H_{{\cal F}}-H_{S{\cal F}},
\end{equation}
where $H$ is the von Neumann entropy.
The emergence of Darwinism can be explained by the following example: the pure
 system+environment state 
\begin{equation}
|\phi_{SE}\rangle=(\alpha| 0\rangle +\beta | 1\rangle )\otimes |
\varepsilon^{(0)}, \varepsilon^{(1)},\dots, \varepsilon^{(N)} \rangle
\end{equation}
evolves into 
\begin{equation}
|\phi_{SE}^{\prime}\rangle=\alpha| 0\rangle | \varepsilon_0^{(0)},
\varepsilon_0^{(1)},\dots, \varepsilon_0^{(N)} \rangle +\beta | 1\rangle  |
\varepsilon_1^{(0)}, \varepsilon_1^{(1)},\dots, \varepsilon_1^{(N)} \rangle,
\label{ghz}
\end{equation}
where $\langle \varepsilon_i^{(n)}| \varepsilon_j^{(m)}\rangle= \delta_{ij}$.
In such a GHZ state, the information carried out by the system can be also found
considering any possible environment fraction of any size. In fact, all the
reduced density matrices have the same entropy and, for any $\#\mathcal{F}$ excluding $0$ and $N$, $I(S:{\mathcal F})= H_S$, where $\rho_S$ is the density matrix of the
system after the bath has been traced out.
If the total state $S+{\cal E}$ is pure, measuring the whole environment, one
acquires full knowledge about the state of the system, given that $H_S=H_{{\cal
E}}$.
As a further example, let us consider a random pure state. Is it  possible to get a high
amount of information monitoring only a small part of the bath? What happens is
that, typically, the observer cannot learn anything about a system without
sampling at least half of its environment. This characteristic behavior is illustrated
in Refs.\cite{zurekabstract1,zurekabstract2}. In turn, states created by decoherence, which do not
follow this behavior  and are those explaining our everyday classical experience, have zero measure in the thermodynamic limit.

It is pretty natural to ask whether the presence of correlations in the
environment facilitates the emergence of quantum Darwinism. Here, we address this 
problem considering the following model: a single spin $S$ (the system) couples 
to a collection of $N$  spins, decoupled from each other. The Hamiltonian 
considered  is $H=H_S+H_B+H_{S\cal E}$, where
\begin{eqnarray}
H_{B}&=&-\sum_{i=1}^N (\sigma_i^+\sigma_{i+1}^- +
\sigma_i^-\sigma_{i+1}^+)-h\sum_{i=1}^N \sigma_i^z,\label{hb}\\
H_{S\cal E}&=& d\; \sigma_S^z\otimes\sum_{i=1}^N \sigma_i^x.
\end{eqnarray}
Here, $N$ is the total number of spins in the bath, $h$ is the strength of the external magnetic field, and periodic boundary
conditions are imposed.
Assuming $[H_S,H_{S\cal E}]=0$, the system timescales are of no importance and will
be neglected, that is, we will neglect $H_S$ unless otherwise specified.

 Usually, open quantum systems are studied in the weak-coupling regime. However,
as we have said in the introduction, we want to investigate the ability of this
spin bath to store classical copies of the system. Then, the bath itself must be
modified during the dynamics.  To this end, we will consider a system-bath
interaction much higher than the bath Hamiltonian itself ($d\gg h\sim 1$),
whose effects on the dynamics can be neglected at least as a first
approximation.

The initial state of the system is
$|+\rangle=(|\uparrow\rangle+|\downarrow\rangle)/\sqrt{2}$, while, assuming zero
temperature, the bath is initially prepared in its  ground state $\ket{G}$. As
$[H_S, H_{S\cal E}]=0$, the evolution of the global state takes the form
\begin{equation}
|\phi_{SE}(t)\rangle=\frac{1}{\sqrt{2}}(|\uparrow\rangle
\ket{G^\uparrow}(t)+|\downarrow\rangle \ket{G^\downarrow}(t) ).
\end{equation}
The system undergoes pure dephasing, that is, its density matrix evolves in time
as
\begin{equation}
\rho_S(t)=\frac{1}{2}\left(
\begin{array}{cc}
1 & \nu(t)\\ \nu^*(t) & 1
\end{array}\right),
\end{equation}
where $\nu(t)=\langle G^\downarrow \ket{G^\uparrow}$.

As we are going to prove,  different forms for the ground state show different
capability of storing redundant copies of classical information of the system.

\section{Quantum phases}\label{I}

In this section we briefly review some of the properties of the isotropic $XX$
Hamiltonian in the presence of a transverse field, which is exactly the  model
introduced in  \eqref{hb}
 to describe the bath.

The exact spectrum of $H_B$  can be calculated using the Jordan-Wigner
transformation, which maps spins into spinless fermions \cite{lieb}. 
This model is known not to possess a proper quantum phase transition (QPT).
Instead, an infinite-order Kosterlitz-Thouless (KT) \cite{kt} quantum phase transition
without symmetry breaking takes place around $h=h_C=1$.

Depending on the value of the transverse field $h$, which plays the role of an
effective chemical potential, the number of fermionic excitations in the ground
state changes. In the following, $\ket{G_n}$ will indicate the ground state in
the $n$-excitation sector. Notice that each $\ket{G_n}$ is the ground state for
a finite range of values of $h$, and that
the number of sectors $\bar n$ depends on the number of spins in the bath
as $\bar n=(N/2+1)$ for $N$ even and $\bar n=(N+1)/2$ for $N$ odd. For $h\ge
h_C$, no excitations are present, that is, the ground state is $\ket{G_0}$,
while $\ket{G_{\bar n-1}}$ is the ground state in the region near $h=0$. In the
thermodynamic limit, the magnetization grows continuously from $h=0$ up to $h=1$
and then remains constant showing a cusp around $h=h_C$ \cite{barouch}. 

\section{Results}

\subsection{Quantum Darwinism from ground state ordering}\label{exact}
As  said before, we  discuss the emergence of Darwinism preparing the bath in its
ground level. At the initial time we assume
\begin{equation}
|\psi(0)\rangle= \ket{+}\otimes \ket{G_n}. \label{psizero}
\end{equation}
 In the strong coupling regime, that is, neglecting $H_B$ during the dynamics,
the evolution is given as
\begin{equation}
|\psi(t)\rangle=e^{-i H_{S\cal E}t}|\psi(0)\rangle. \label{psit}
\end{equation}

Because of the form of the interaction, the initial state evolves spanning the
whole set of $2^{N+1}$ basis states. Such an exponential dependence strongly
limits the maximum number of spins that can be introduced to obtain the
 numerical solution in computationally reasonable times. However, if the
bath's initial (ground) state is either $|G\rangle=|G_0\rangle$ or
$|G\rangle=|G_1\rangle$, due to the invariance of these states under any
spin-spin swap,  the effective action of $H_{S\cal E}$  takes a very simple form. Let
us indicate with $|\textbf{i}\rangle$ the $i$-magnon state, that is the
swap-invariant state with $i$ $1$s and $(N-i)$ $0$s; for instance
$|G_0\rangle=|\textbf{0}\rangle$ and $|G_1\rangle=|\textbf{1}\rangle$.
Consider also that 
\begin{eqnarray}
e^{-i H_{S\cal E}t}|\psi(0)\rangle&=&e^{-i
d\sigma_S^z\sum_i\sigma_i^xt}|\psi(0)\rangle\nonumber\\
&=&e^{-i d\sum_i\sigma_i^xt}|\uparrow\rangle|G\rangle+e^{+i
d\sum_i\sigma_i^xt}|\downarrow\rangle|G\rangle\nonumber.
\end{eqnarray}
 It is easy to show that $\sum_i\sigma_i^x$ couples $|\textbf{i}\rangle$
to $|\textbf{i+1}\rangle$ and to $|\textbf{i-1}\rangle$, with no other states
involved.

Then, as detailed in the appendix, an effective Hamiltonian with $N+1$
degrees of freedom can be built to describe evolution \eqref{psit} of \eqref{psizero} :
\begin{equation}
H_{{\rm eff}}=\sum_{n=0}^{N}(A_{n}^+|\textbf{n}\rangle\langle
\textbf{n+1}|+A_{n}^-|\textbf{n}\rangle\langle\textbf{n-1}|)
\end{equation}
where $A_{n}^-=\sqrt{n(N-n+1)}$ and $A_{n}^+=A_{n+1}^-$. As the transition from
$|G\rangle=|G_0\rangle$ to $|G\rangle=|G_1\rangle$ takes place at $h_C$, we can
monitor the qualitative change around the critical point by considering very
long chains.

The simple form of the evolution operator allows us to learn about the
recurrence time of the system. In fact,  we have
\begin{equation}
|\psi(\pi/2d)\rangle=U_X  |\psi(0)\rangle,
\end{equation}
where $U_X=\otimes_{i=1}^N \sigma^x_i$ is the spin flip operator. Then, at
$d\cdot t=\pi/2$, the evolved state is identical to the initial one
provided that the exchange $|\uparrow\rangle \leftrightarrows |\downarrow\rangle
$  has been applied to every single spin. Then, as far as the information
content of the state is considered,  $\pi/2$ represents the model periodicity,
and  $\pi/4$ the time when the  influence of the environment over the state is
maximized before revival takes place. Based on these considerations, we also
expect that quantum Darwinism effects are more evident for  $\pi/4$.

\begin{figure}[t]
\includegraphics[width=9cm]{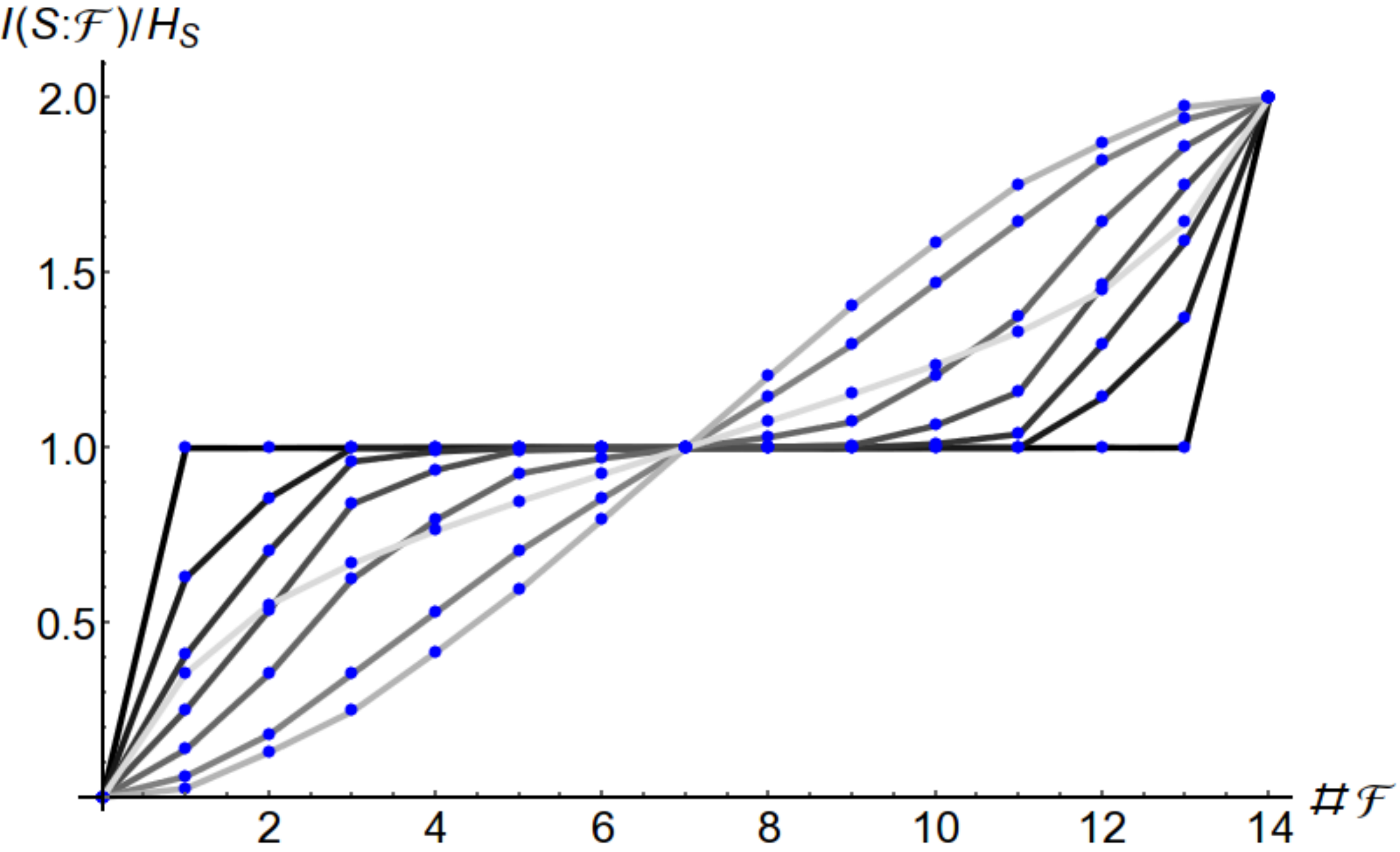}
\caption{${\cal I}(S:{\mathcal F} )$ at $t=\pi/4$ for $N=14$ and in the
strong-coupling regime $d=\infty$ for different values of $h$. The eight lines
correspond to the eight different ground states of $H_B$. The darker the line,
the higher the value of $h$. For $h=h_C=1$ (black curve) ${\cal I}(S:{\mathcal F} )=H_S$ 
for any size of the fraction: even observing only one spin in the bath
we can learn all the classical information on the system; quantum Darwinism has been fully achieved.
For lower $h$   (the lightest line corresponds to the ground state of the sector $0\leq h\lesssim 0.11$) a significant fraction size ($\#{\cal F}\sim N/ 2$) has to be observed in order to reach enough
information (${\cal I}(S:{\mathcal F} )\sim H_S$), so quantum Darwinism has not succeeded.}
\label{fig1}
\end{figure}

In Fig. \ref{fig1}, we chose the optimal time $d\cdot t=\pi/4$ and calculated
${\cal I}(S:{\mathcal F} )$ as a function of $\#\mathcal{F}$ and for different
values of $h$ in order to take into account any of the possible ground states
$\ket{G_n}$ of $H_{B}$ for a bath of $14$ spins. From Fig. \ref{fig1}, we learn
that the presence of correlations in the initial state ($h<h_C$) has the tendency to
destroy the emergence of quantum Darwinism, while as expected, uncorrelated states
($h\geq h_C$) bring perfect redundant information proliferation. When moving towards
lower values of $h$ (lighter curves) the behavior of ${\cal I}(S:{\mathcal F} )$ tends to the one 
observed picking random states (no Darwinism); however for very low $h$ the shape
returns to a Darwinism-like one, even though the achieved slope is far from being
optimal. 

A possible intuitive explanation of the observed trend, although it will have to
remain as a conjecture at this stage, comes from studying the entanglement
properties
of ground states of the XY model, both bipartite \cite{osborne} and multipartite
\cite{giampaolo}.
When we take a given fraction, its initial entropy is higher when it is more entangled with the rest of the bath, thus more (multipartite) entanglement means more initial entropy of
the fraction and thus less ability to store new info about the system. Thus, as discussed in \cite{haziness}, the best situation for Darwinism is when each bath unit 
is initially pure (no entropy, so it can grow its entropy maximally through 
learning the system's state) and when they are orthogonally aligned to the eigenbasis of the interaction Hamiltonian (this is why the $h=h_C$ case is optimal).
As we go away from $h=h_C$ to lower field values, multipartite entanglement increases each fraction entropy and worsens the ability to produce darwinism.
However, when the ground state is half filled, that is, when $h\sim 0$, a new competing effect appears: the system von Neumann entropy drops dramatically close to zero at the time when 
Darwinism is expected to appear (not shown).  At the same time, the bath's single fraction density matrix remains maximally mixed. Because of this drop, the difference between 
$H_{{\cal F}}$ and $H_{S{\cal F}}$ becomes less relevant and the Darwinism measure is more sustained (i.e. better) than in the presence of more intense values of the magnetic field. 
This means that we are facing a non-linear type effect: the multipartite entanglement initially present in the bath is able to induce a small amount of disorder in the 
system without losing its properties.

\subsection{Mixing of classical records}
We have seen that initially correlated states in the bath ($h<h_C=1$) perform
worse in terms of quantum Darwinism.  This means that they are less able to store
redundant copies of the classical information
about the system. A further aspect, as studied e.g. in \cite{zurekFALL}, is
whether classical redundant copies are stable along time. In that work, the
authors study a spin dephasing star model in which they add
random weak couplings among bath units; such couplings introduce mixing among
the records, thereby leading to a delocalization of the system's information. It
is thus no longer possible to recover
such information by just measuring a small, local, fraction of the environment,
hence leading to poor Darwinism.
\begin{figure}[h!]
\includegraphics[width=7cm]{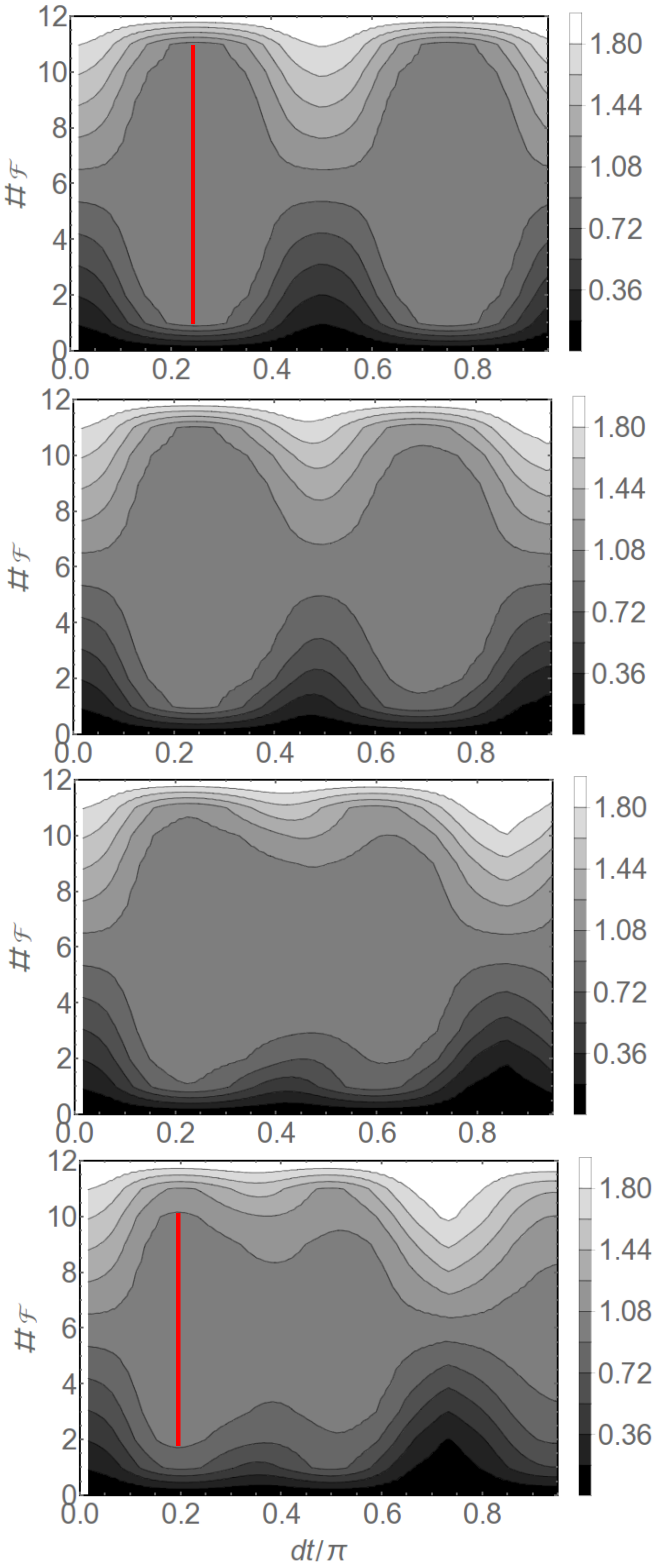}
\caption{Full evolution of ${\cal I}(S:{\mathcal F} )$ including the bath's
Hamiltonian, with $d=h=1$, $N=12$ and $H=H_{S\cal E}+\lambda H_B$. From top to
bottom: $\lambda=0,0.25,0.5,1$. In the $\lambda=0$ case a perfect plateau is
seen at $d\cdot t=\pi/4$ (we have drawn a red line to highlight the plateau) 
and full dynamics recurrence time occurs at $d\cdot t=\pi/2$. Once the bath Hamiltonian comes into play $\lambda\neq0$, the first
plateau gets narrower due to mixing of records in the environment, and even more
so the second plateau.  Also the recurrence dynamics gets shorter and distorted.
The more important is the role of the bath's Hamiltonian, for higher $\lambda$,
the more pronounced is the effect. Mixing of bath records breaks the quantum
Darwinism plateau. }
\label{fig2}
\end{figure}

Here we have so far studied the case of evolution dictated by only $H_{S\cal E}$,
which imprints the information of the system onto the bath. But what happens if
the self-dynamics of the bath is taken into account?
The Hamiltonian $H_B$ consists of a local (field $h$) part and a term (XX
interaction) which propagates interaction among bath units, therefore it is to
be assumed that $H_B$ will mix different local records
and diffuse such information in a nonlocal fashion all over the bath's
extension. If so, a worse plateau should be observed.
Indeed, in Fig. \ref{fig2} we see that such is the case. There we let system and
bath evolve according to $H=H_{S\cal E}+\lambda H_B$ for different values of
$\lambda$ (to be compared with results in previous 
section where we set $d=1$ and $\lambda=0$, which is the equivalent of $d\gg1$).
A progressive worsening of the plateaus for higher $\lambda$s can be observed,
in addition to a distortion of the periodicity which was
present when only $H_{S\cal E}$ was active.

\subsection{Non-Markovianity}
So far, we have studied the emergence of Darwinism induced by the system-bath
interaction. As we are working in the strong-coupling regime, it is quite
natural to expect that memory effects  come out together with a a flux of
information from the bath to the system. 
The connection between  non-Markovianity and Darwinism was discussed in Ref.
\cite{galve14} considering a harmonic oscillator whose position was coupled to
the positions of $N$ harmonic oscillators representing the bath. In that
context, it was shown that the presence of memory effects inhibits the emergence
of objective reality. The same kind of analysis can be carried out considering
the model under investigation.

The measure introduced in Ref. \cite{nm} provides a way of quantifying the
amount of non-Markovianity produced during the dynamics. Let us briefly recall
such a definition.
In a Markovian process, the distinguishability between any pair of quantum
states is a monotonously decreasing function of time. Then, the presence of time
windows where some states become more distinguishable between each other
witnesses the presence of non-Markovianity in the dynamical map. The
non-Markovian quantifier is then defined as
\begin{equation}
 \label{eq:N}
\mathcal{N}=\max_{\rho_1,\rho_2}\int_{\sigma>0}dt\;\sigma(t,\rho_{1,2}(0)),
\end{equation}
where the trace distance, which quantifies distinguishability is
$D(\rho_{1},\rho_{2})={\rm Tr}|\rho_{1}-\rho_{2}|/2$ and 
where its rate of change  is
$\sigma[t,\rho_{1,2}(0)]=dD[\rho_{1}(t),\rho_{2}(t)]/dt$. The maximum in Eq.
\eqref{eq:N} is taken over any possible pair of states $\{\rho_1,\rho_2\}$. In
the case of pure dephasing, it was shown in Ref. \cite{he} that  there is a simple
way of calculating $\mathcal{N}$ in terms of the Loschmidt echo: 
the trace distance of any pair of (qubit) system states is
$D[\rho^{1}_{s}(t),\rho^{2}_{s}(t)]=\sqrt{L(t)}$, where the Loschmidt echo is
$L(t)=|\nu(t)|^2$.

\begin{figure}[t]
\includegraphics[width=8cm]{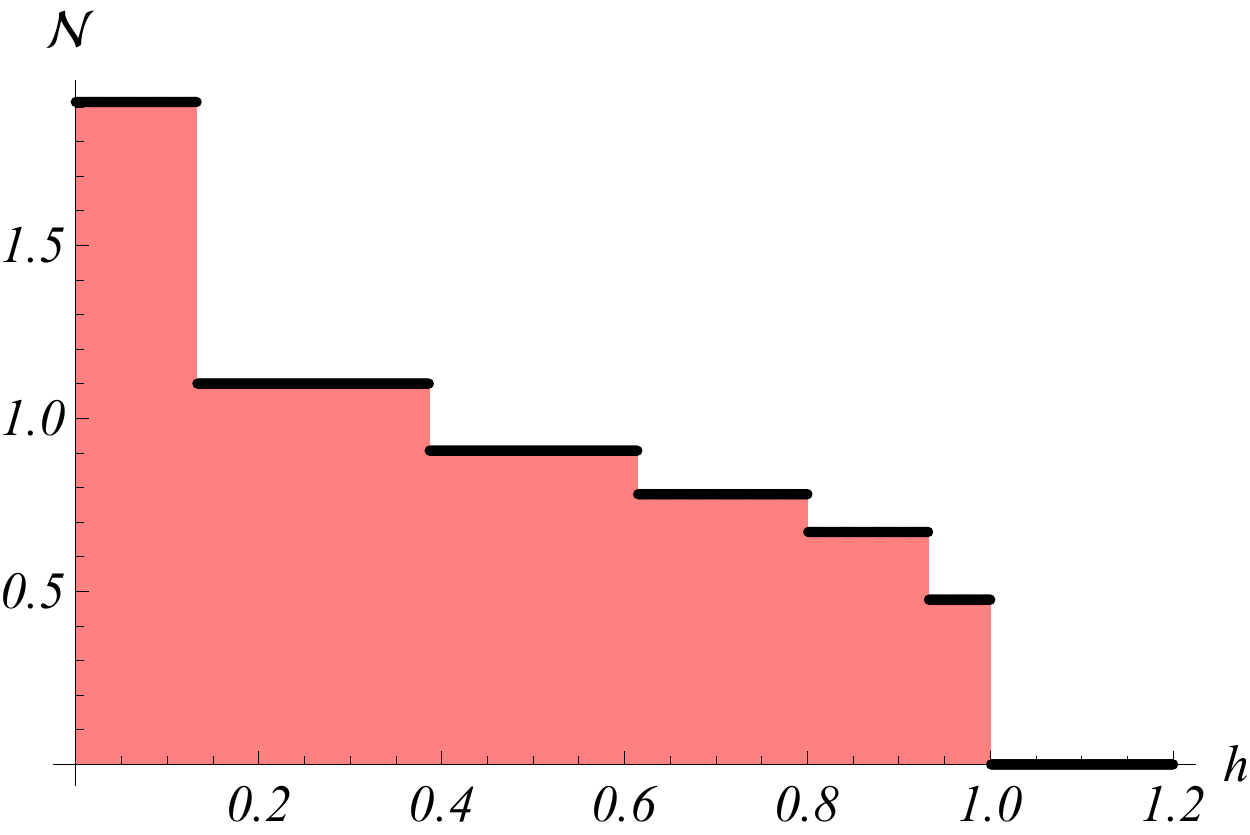}
\caption{Non-Markovian quantifier $\mathcal{N}$ for a $12$-spin bath in the
strong-coupling regime.  }
\label{fignm1}
\end{figure}

In Fig. \ref{fignm1} we plot ${\cal N}$ as a function of $h$ for  a bath of 
$12$ spins. The measure is taken considering times between zero and $\pi/4$. In
agreement with the results of Ref. \cite{galve14}, the evolution is completely
Markovian for $h\ge h_C$. By lowering the value of $h$ and passing through the
whole family of $\ket{G_n}$, the value of  ${\cal N}$ increases monotonically as
magnetization decreases and reaches its maximum for $h=0$. From this point of
view, the result is qualitatively similar to the one obtained considering
Darwinism. In other words, the higher the amount of non-Markovianity, the smaller
the ability of proliferating throughout the environment.

\begin{figure}[t]
\includegraphics[width=8cm]{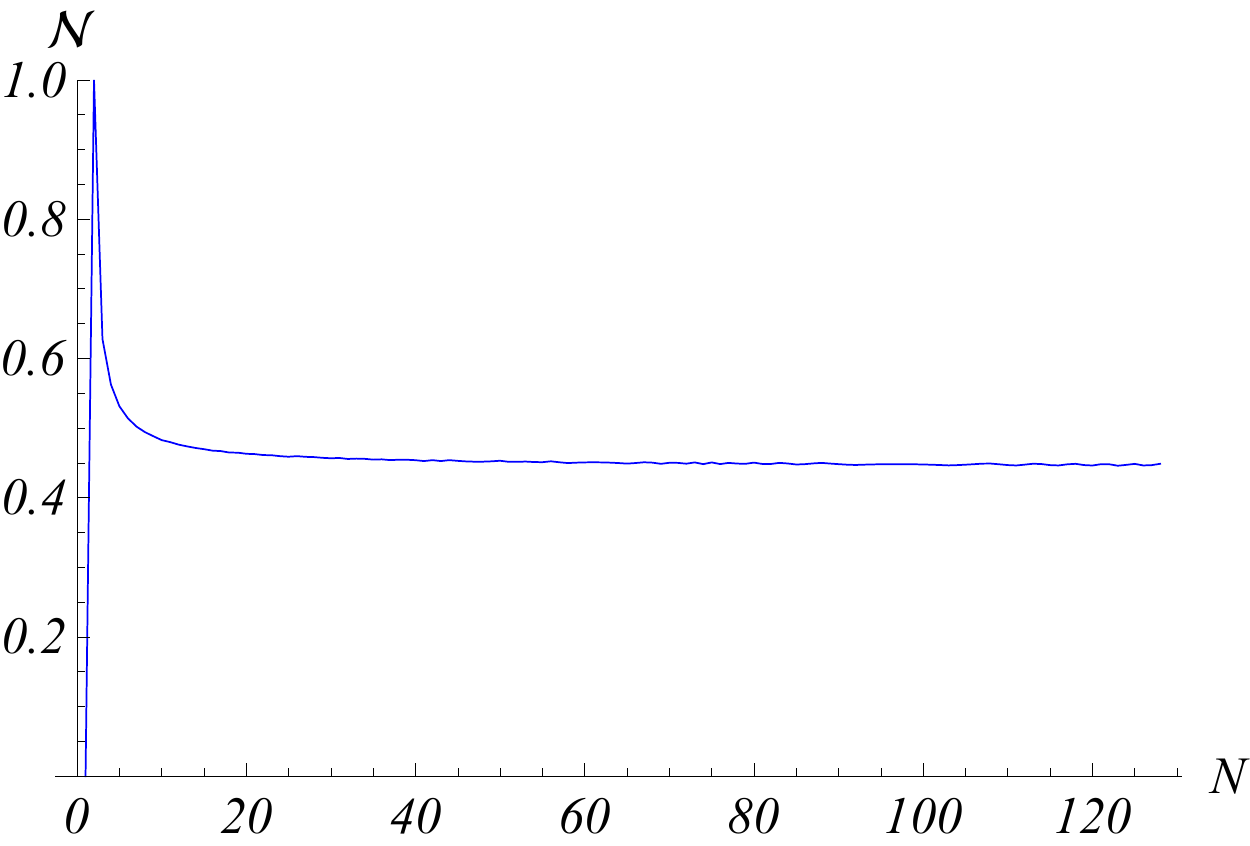}
\caption{Non-Markovianity immediately before the critical point
$\mathcal{N}(h_C^-)$ as a function of the bath size $N$.  }
\label{fignm2}
\end{figure}

A more detailed analysis can be given calculating  $\mathcal{N}$ around $h=h_C$
by means of the exact solution introduced in Sec. \ref{exact}. As expected, for
$h\ge h_C$, $\mathcal{N}=0$ irrespective of the bath's size. More interesting is
the behavior immediately below the critical point.
Indeed, except for the case of very
short chains ($N\lesssim 8$), $\mathcal{N}(h_C^-)$ has a constant value (see
Fig. \ref{fignm2}). In a way,  $\mathcal{N}$  plays the role of a precursor of
KT phase transition, as the critical point can be spotted far before reaching
the thermodynamic limit by monitoring the amount of non-Markovianity. 

\section{Conclusions}

In conclusion, we have studied several aspects of the emergence of quantum 
Darwinism in the pure-decoherence setting due to a spin bath. We have chosen 
to initialize the bath in different quantum phases of the isotropic XX  model 
with transverse field. If the initial bath's state is uncorrelated ($h\ge h_C$), as already known,
quantum Darwinism arises as redundant  proliferation of information is produced
throughout the bath. The presence of initial correlations ($h<h_C$), however, 
represents an obstacle towards the building up of classical objectivity. 

We have also quantified the amount of non-Markovianity of the system's
dynamics, finding that it correlates well with the absence of quantum Darwinism.
This offers further evidence of what was already shown in \cite{galve14} for the Brownian
oscillator model.

Finally, along the lines presented in \cite{zurekFALL}, we have shown that coupling
between spins in the bath necessarily ruins the achieved redundant classical information
storage. This can be understood as mixing and delocalization of such records, whereby
information can no longer be gained locally through small bath's fragments.

As a main conclusion, we have found that the quantum Darwinism program is better achieved if
the environment is similar to a blank slate (no correlations) made of uncoupled units, as
intuitively is to be expected from any good memory device.

\acknowledgments
 F.G. and R. Z. acknowledge funding from  MINECO, CSIC, EU commission, FEDER under Grants No.
FIS2007-60327 (FISICOS) and No. FIS2011-23526 (TIQS), postdoctoral JAE program (ESF), and COST Action MP1209.
 G.L. G. acknowledges funding from Compagnia di San Paolo.

\appendix*

\section{Simplified exact solution around the critical point}

According to the notation introduced in the main text, for fields immediately
smaller than $h_C$, the ground state of $H_B$ is the single-magnon state 
\begin{equation}
|G_1\rangle=\frac{1}{\sqrt{N}}\sum_i |0,0,\dots,1_i,\dots,0\rangle.
\end{equation}

Let us assume that the bath is initially prepared in $|G\rangle=|G_0\rangle$ or
in $|G\rangle=|G_1\rangle$, while the initial state of the system is
$|+\rangle=(|0\rangle+|1\rangle)/\sqrt{2}$.
Due to the purely dephasing character of the coupling, the evolution in time
$|\psi(t)\rangle=e^{-i H t}|G\rangle\otimes  |+\rangle$
can be split into 
\begin{equation}
|\psi(t)\rangle=\frac{1}{\sqrt{2}}(e^{-i H_0 t}|G\rangle\otimes  |0\rangle+e^{-i
H_1 t}|G\rangle\otimes  |1\rangle)
\end{equation}
where $H_0=-H_1=\sum_{i=1}^N \sigma_i^x$ only involve bath degrees of freedom.

Given that  $|G\rangle$  is invariant under any spin-spin swap, its evolution
takes a very simple form. Let us indicate with $|\textbf{i}\rangle$ the
$i$-magnon state, that is the swap-invariant state with $i$ $1$s (let us point
out that $|\textbf{i}\rangle\neq |G_i\rangle$ unless $i=0,1$).  It is easy to
show that $H_0$ couples $|\textbf{i}\rangle$ to $|\textbf{i+1}\rangle$ and to
$|\textbf{i-1}\rangle$, with no other states involved. Then, an effective
Hamiltonian with $N+1$ degrees of freedom can be built to describe the evolution
of $|G\rangle$:
\begin{equation}
H_{{\rm eff}}=\sum_{n=0}^{N}(A_{n}^+|\textbf{n}\rangle\langle
\textbf{n+1}|+A_{n}^-|\textbf{n}\rangle\langle\textbf{n-1}|)
\end{equation}
where $A_{n}^-=\sqrt{n(N-n+1)}$ and $A_{n}^+=A_{n+1}^-$.

Once the effective Hamiltonian has been derived, the exact dynamics of the
initial state $|G\rangle\otimes |+\rangle$ takes the form
\begin{equation}
|\psi(t)\rangle=\frac{1}{\sqrt{2}}\sum_{n=0}^N [c_n(t)|\textbf{n}\rangle\otimes
|0\rangle+c_n(-t)|\textbf{n}\rangle\otimes |1\rangle].
\end{equation}
The reduced density matrix of the system is then given by
\begin{equation}
\rho_S(t)=
\frac{1}{2}
\left(\begin{array}{cc}
1&\nu(t)\\
\nu^*(t)&1
\end{array}\right)
\end{equation} 
where $\nu(t)=\sum_n c_n^2(t)$.

If we want to calculate the reduced density matrices of fractions of the bath we
need to decompose the states $\textbf{n}$ into sub-parts.
As all the states involved have long-range correlations, we will limit our
calculation to considering the first $k$ spins of the baths (the result would be
identical for any group of $k$ spins). So,  defining the partition $\{k,N-k\}$,
the state $|\textbf{n}\rangle_N$ (the subscript indicates the Hilbert space
where the state is defined) can be written as
\begin{equation}
|\textbf{n}\rangle_N=\sum_{i=i_{\min}}^{i_{\max}}f_{N,n,i,k}|\textbf{i}
\rangle_k|\textbf{n-i}\rangle_{N-k}
\end{equation}
where $i_{\max}=\min[k,n]$, $i_{\min}=\max[0,k+n-N]$ and where
\begin{equation}
f_{N,n,i,k}=\sqrt{{N-k \choose n-i}{k \choose i}/{N \choose n}}.
\end{equation}

Eliminating $k$ spins from the bath gives\begin{widetext}
\begin{eqnarray}
\rho_{\bar k,S}(t)&=&\frac{1}{2} \sum_{i=0}^k\sum_{n,m=0}^{N-k}
c_{n+i}(t)c_{m+i}^*(t)f_{N,n+i,i,k}f^*_{N,m+i,i,k}| \textbf{n}\rangle\langle
\textbf{m}|\otimes|0\rangle \langle 0|\nonumber\\
&+&\frac{1}{2} \sum_{i=0}^k\sum_{n,m=0}^{N-k}
c_{n+i}(t)c_{m+i}^*(-t)f_{N,n+i,i,k}f^*_{N,m+i,i,k}| \textbf{n}\rangle\langle
\textbf{m}|\otimes|0\rangle \langle 1|\nonumber\\
&+&\frac{1}{2} \sum_{i=0}^k\sum_{n,m=0}^{N-k}
c_{n+i}(-t)c_{m+i}^*(t)f_{N,n+i,i,k}f^*_{N,m+i,i,k}| \textbf{n}\rangle\langle
\textbf{m}|\otimes|1\rangle \langle 0|\nonumber\\
&+&\frac{1}{2} \sum_{i=0}^k\sum_{n,m=0}^{N-k}
c_{n+i}(-t)c_{m+i}^*(-t)f_{N,n+i,i,k}f^*_{N,m+i,i,k}| \textbf{n}\rangle\langle
\textbf{m}|\otimes|1\rangle \langle 1|
\end{eqnarray}
\end{widetext}

Let us assume that the bath is initially prepared in $|\textbf{n=0}\rangle$. For
$t=\pi/4$, the state takes the form
\begin{equation}
|\psi(t)\rangle=\frac{1}{\sqrt{2}} [(|0\rangle+i|1\rangle)^{\otimes N}\otimes
|0\rangle+[(|0\rangle-i|1\rangle)^{\otimes N}\otimes |1\rangle]
\end{equation}
and brings perfect Darwinism.


\end{document}